\documentclass[twocolumn,showpacs,preprintnumbers,amsmath,amssymb,prb,floatfix]{revtex4}

\usepackage{graphicx}
\usepackage{dcolumn}
\usepackage{bm}
\usepackage{epstopdf}
\begin{document}

\title{Magnetic and Orbital Order in LaMnO$_3$ under Uniaxial Strain: A Model Study
}

\author{B. R. K. Nanda and S. Satpathy}  
\affiliation{Department of Physics, University of Missouri, Columbia, Missouri 65211, USA  
}
\date{\today}

\begin{abstract}
The effect of uniaxial strain on electronic 
structure and magnetism in LaMnO$_3$ is studied from a  
model Hamiltonian that illustrates the competition between the Jahn-Teller, super exchange, and double 
exchange interactions. We retain in our model the three main octahedral distortions ($Q_1,  Q_2$, and 
$Q_3$), which couple to the Mn $(e_g)$ electrons. 
Our results show the ground state to be a type A 
antiferromagnetic (AFM) insulating state for the unstrained case, consistent with experiments. With tensile strain (stretching along the c axis), the ground state
 changes into a ferromagnetic and 
eventually into a type G$^\prime$ AFM structure, while with compressive
strain, we find the type A switching into a type G structure. 
The orbital ordering, which displays the well known checkerboard 
$x^2-1 / y^2-1$ structure for the unstrained case, retains more or less the same character for compressive strain, while changing into the $z^2-1$ character for tensile strains. 
While $Q_1$ and $Q_3$ are fixed by the strain components 
$\varepsilon_{xx}$ and $\varepsilon_{zz}$ in our model, the magnitude of the in-plane 
distortion mode (Q$_2$), which varies to minimize the total energy, slowly diminishes with tensile strain, 
completely disappearing as the FM state is entered. Within our model, 
the FM state is metallic, while the three AFM states are  
insulating.

\end{abstract}

\maketitle
\section{Introduction}

It is well known that the interplay between the orbital, lattice, and spin degrees of freedom in the lanthanum manganite, a key member of the colossal magnetoresistive materials, produces a checkerboard orbital-ordered ground state by the cooperative Jahn-Teller effect and simultaneously the type A antiferromagnetic ordering. However, a detail knowledge of what happens to this ground state under the application of strain or uniform pressure is lacking. 

On the other hand, high-quality, lattice-matched heterostructures of the manganites have begun to be grown and studied, heterostructures that are frequently under uniaxial strain due to the pseudomorphic growth condition during epitaxial growth. The 
variation of layer thickness and substrate induce different  
magnetic ground states ranging from complete ferromagnetic ordering 
to a canted magnetic state or an antiferromagnetic state\cite{christides, yamada, 
koida, satoh, smadici}. An interesting two-dimensional spin-polarized electron gas was also predicted for the manganite heterostructures.\cite{brk3} In addition, the magnetotransport
studies on bulk LaMnO$_3$(LMO) under pressure \cite{ramos, loa, gaudart, ding}  
and LMO thin films grown on different substrates
\cite{tebano, konishi, zhang, song, aditi, suzuki, sun} show metal 
insulator transitions accompanied by magnetic transitions.  Theoretically, it is known that strain can change orbital ordering in LMO, which affects magnetism,\cite{meskine} and leads to modified properties in the manganite superlattices.\cite{brk,brk2} The commonalities involved in these are strain, 
either induced by pressure or due to substrate.  

It is therefore important to understand the ground-state of the underlying material, viz., the bulk LMO under uniaxial strain condition.  It is the goal of the paper to provide this insight by studying a simple model Hamiltonian that contains the relevant interactions important for this material. 

\begin{figure}
\includegraphics[width=7cm]{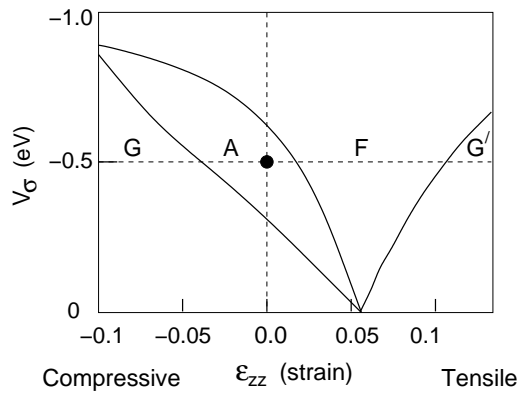}
\caption{\label{magneticphase} The ground-state magnetic phase diagram for LMO as a function of strain and  the $dd\sigma$ electronic hopping $V_\sigma$, as obtained from our model.  Type A represents the phase with ferromagnetic $ab$ planes stacked antiferromagnetically along the $c$ axis, while F and G stand, respectively,  for the ferromagnetic and the Ne\'{e}l ordered antiferromagnetic state. We distinguish between two
type G states (G and G$^\prime$) depending on the orbital occupancy of the Mn $e_g$ electron: $x^2-1$ or $y^2-1$ for G and $z^2-1$ for G$^\prime$.  For LMO, $V_\sigma$ is about -0.5 eV. 
}
\end{figure}  
%

\begin{figure*}
\includegraphics[width=13cm]{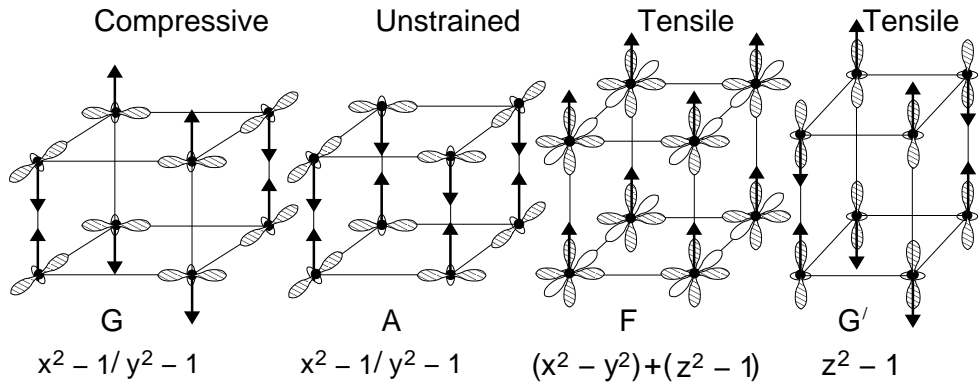}
\caption{\label{orbital} The magnetic and orbital order of the four  phases under varying strain conditions as obtained from our model. Shown are only the Mn atoms and arrows denote the $t_{2g}$ core spins.
Tensile strain means stretching along the $c$ direction, normal to the $ab$ plane.}
\end{figure*} 

Density-functional studies\cite{satpathy,singh,pickett,zenia} of the unstrained LMO have 
provided important insights into the 
physics of the system regarding how the Jahn-Teller (JT) coupling, orbital 
ordering, and magnetism are intertwined. The few such studies performed  for the strained LMO are focused on the uniform  pressure situation and indeed they do find changes in the magnitude of the JT distortions as well as in the conduction properties.\cite{trimarchi, yamasaki} Above the critical pressure of  32 GPa, LMO becomes metallic and the JT distortion appears to be completely suppressed.\cite{loa, gaudart} Systematic experimental work as well as theory work on the uniaxial strain situation is missing;  yet, as already mentioned, this strain condition is frequently realized  in thin films and superlattices grown on different substrates and must be understood before an understanding of the superlattices of LMO can be accomplished. 
 
In this paper, we study the interplay between the lattice, orbital, and magnetic degrees of freedom in  LMO under uniaxial strain condition by using a tight-binding model that includes the relevant interactions, viz., 
the antiferromagnetic super exchange between the t$_{2g}$ core spins, 
the double exchange between the core spins and the itinerant e$_g$ electrons, the electronic band structure energy, as well as the  JT coupling
between the e$_g$ electrons and the lattice. 

At this stage, a model study is more instructional than a density-functional study, since unlike the latter,  one can vary here the relative strengths of the various interactions and study their effect on the ground state, providing us with valuable insight into the system. This study will be important in understanding the experimental results on strained LMO as well as help interpret results of detail {\it ab initio} calculations, which are in progress and will be reported elsewhere.\cite{Nanda-LMO}
 
Our results, summarized in Figs.  \ref{magneticphase} and \ref{orbital},  show that the magnetic state, the JT distortions, the orbital ordering, 
as well as conduction properties, all change with uniaxial strain.
As strain changes from  compressive to tensile, the system changes from a type G antiferromagnetic (AFM)  
to the type A AFM phase, then to a ferromagnetic phase (F), ultimately changing into a type G$^\prime$ phase. The FM phase is metallic, while the rest are insulating phases. As regards the orbital ordering, while  the unstrained LMO shows the well known checkerboard 
$x^2 - 1$/$y^2-1$ ordering on the $ab$ plane, the application of compressive strain retains the same orbital structure more or less, while
tensile strain results in a $z^2 - 1$ orbital ordering. As the strain is changed from compressive to tensile, the magnitude of the in-plane distortion ($Q_2$ mode) diminishes, eventually disappearing completely for the ferromagnetic phase and beyond. The changes in the various features are intertwined with one another
as discussed below.

\section{The Model Hamiltonian}
\begin{figure}
\includegraphics[width=8cm]{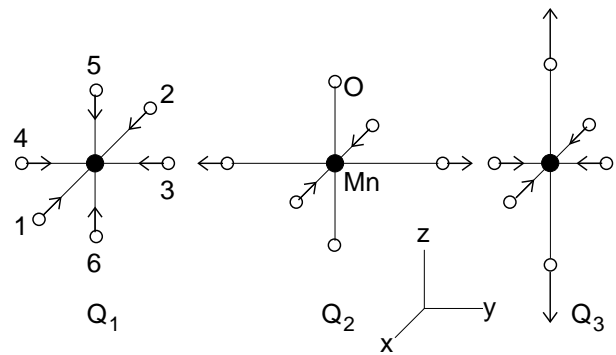}
\caption{\label{q1q2q3} Vibronic modes 
for the MnO$_6$ octahedron with the eigenvectors: 
$|Q_1\rangle = (-X_1 + X_2 - Y_3 + Y_4 - Z_5 + Z_6)/\sqrt{6}$, 
$|Q_2\rangle = (-X_1 + X_2 + Y_3 - Y_4)/2$, and 
$|Q_3\rangle = (-X_1 + X_2 - Y_3 + Y_4 + 2Z_5 - 2Z_6)/\sqrt{12}$,
where $X_1$ denotes the $x$ coordinate of the first oxygen atom, etc.}
\end{figure}

We consider the following model Hamiltonian, restricted to the Mn sites in a tetragonal lattice, with lattice constants being different along the plane and normal to the plane in order to accommodate the uniaxial strain:  
 \begin{equation}
{\cal H} = \sum_{i}\frac{1}{2}KQ{_i}^2 + \sum_i {\cal H}_{JT}^i + {\cal H}_{ke} + 
\frac{J}{2}\sum_{<ij>}\vec{S}_i \cdot \vec{S}_j,
\label{htot}
\end{equation}
where $i$ is the site index and $<ij>$ denotes summation over nearest-neighbor sites. The last term describes the antiferromagnetic superexchange interaction between the Mn $t_{2g}$ spins, treated as classical, which provide the background for the motion of the itinerant Mn $e_g$ electrons.
The first two terms in the Hamiltonian describe the Jahn-Teller interaction of the Mn $e_g$ electrons with the lattice, where the first term is the elastic energy with $Q^2 = Q_1^2 + Q_2^2+Q_3^2$,
with $Q_1$ being the breathing mode, $Q_2$ being the basal plane distortion mode, and $Q_3$ being the
 octahedral stretching mode as shown in Fig. \ref{q1q2q3}.  These are the three important modes that couple to the $e_g$ electrons.
 
 The JT interaction at each Mn site is given by\cite{kanamori}
\begin{eqnarray}
{\cal H}_{JT} &=& g^{\prime}Q_1I - g (Q_3 \tau_z + Q_2 \tau_x )\nonumber \\
&&- G(Q_3^2 - Q_2^2) \tau_z       + 2GQ_2  Q_3 \tau_x,
\label{hjt}
\end{eqnarray}
where both the linear and quadratic vibronic couplings are included. 
The quadratic coupling is necessary to describe the checkerboard orbital ordering on the $ab$ plane.\cite{popovic} In Eq. \ref{hjt},
$\vec{\tau}$ is the pseudospin describing the two e$_g$ orbitals, viz., $|\uparrow\rangle$ = $|z^2 - 1\rangle$ and $|\downarrow\rangle$ = $|x^2 - y^2\rangle$.   
Denoting the creation operators for these two orbitals on the $i$-th site by $c_{i1}^\dagger$ and 
$c_{i2}^\dagger$, respectively, the pseudospin operator at the $i$-th site in Eq. \ref{hjt} is written as:  $\vec{\tau}_i = \sum_{\alpha \beta} c_{i\alpha}^\dagger  \vec{\tau}_{\alpha \beta} c_{i\beta}$, where the greek indices denote the orbitals and $i$ is the site index. The lattice interactions do not depend on the electron spin, which therefore does not appear in Eq. \ref{hjt}.

The kinetic energy term in the Hamiltonian may be written as
\begin{equation}
{\cal H}_{ke} = \sum_{\langle ij \rangle, \alpha\beta}t_{i\alpha, j\beta}\cos(\theta_{ij}/2) c_{i\alpha}^{\dagger}c_{j\beta} + H.c., 
\label{hke}
\end{equation}
where the value of the Hund's energy is taken as $J_H = \infty$. In this limit, the coupling between the core and the itinerant spins, $ {\cal H}_{Hund} = - J_H \sum_{i\alpha}\vec{S_i} \cdot \vec{s}_{i\alpha}$, makes the electron state with the spin anti-aligned with the $t_{2g}$ core spin at that site inaccessible to the system (core $t_{2g}$ spin and the electron spin at the $i$-th site are denoted by $\vec S_i$ and $\vec s_i$ respectively). Thus the itinerant electrons are effectively described as spinless, with only its spin state aligned parallel to the core spin at a particular site being accessible to the system. The electron hopping is diminished by the well known Anderson-Hasegawa $ \cos (\theta_{ij}/2)$  factor, where $\theta_{ij}$ is the angle between the core spins at the two sites.\cite{anderson, zener}

The nearest-neighbor (NN) hopping integral 
$t$ depends on the relative positions
of the two Mn sites.  
For $NN$ along $\hat x$ or $\hat y$,
we have\cite{harrison}
\begin{equation}
t_{\alpha\beta} = 
\left(\begin{array}{cc}    
1 & -\sqrt{3}\\
-\sqrt{3} & 3\\
\end{array} \right) \times \frac{V_{\sigma}}{4}, 
\end{equation}
while along $\hat z$, we have $t_{\alpha\beta} = \delta_{\alpha 1} \delta_{\beta 1} V_{\sigma}$, where $V_{\sigma}$ is the $dd\sigma$ hopping parameter and we have neglected the much smaller $dd\delta$ interaction.
The typical values of the Hamiltonian parameters used in our calculations throughout,
unless otherwise stated, are: K = 9 eV/{\AA}$^2$ following from the optical studies on La$_{0.85}$Sr$_{0.15}$MnO$_3$\cite{millis1}, g = 2.5 eV/{\AA}, G = 1.5 eV/{\AA}$^2$ and V$_{\sigma}$ = -0.5 eV following earlier density-functional results.\cite{popovic} The value of the exchange energy  is taken as  $J = 26.2$ meV,
which is the experimental value for  CaMnO$_3$,\cite{wollan,rushbroke} which has the $t_{2g}^3e_g^0$ electronic configuration. Note that in our model Hamiltonian, g$^\prime$ is an unimportant parameter, as it affects the energies of both the $e_g$ orbitals equally and therefore adds a constant to the total energy.

We have considered the typical magnetic ordering of the core spins found in the manganites, viz., the  A, C, G, and F types of magnetic states. Type C was always found to be of higher energy within our model. To accommodate these structures, the unit cell in the lattice consisted of four Mn atoms.
The total energy was calculated by diagonalizing the Hamiltonian (\ref{htot}) and adding the energies of the occupied states in the Brillouin zone, corresponding to one $e_g$ electron per Mn atom,  for the various strain conditions and  magnetic configurations.

{\it Strain and the JT parameters} --
In our model, since we keep only the three main octahedral modes, the strain components are related to the magnitudes of the JT distortions. We consider only uniaxial strain conditions but no shear, so that only the diagonal components of the strain tensor are non zero.

For the unstrained crystal, the ground-state in our model has a tetragonal lattice structure with type A magnetic configuration and a checkerboard JT distortion and orbital ordering. The JT distortion has been shown\cite{popovic} to lead to the checkerboard distortions on the $ab$ plane of the type ($Q_2$, $Q_3$) alternating with (-$Q_2$, $Q_3$) consistent with the experiments. 

This checkerboard JT distortion in our model results in a tetragonal structure with the lattice constants being $ a = \sqrt {2/3} \ Q_{10} - Q_{30}/\sqrt{3}$ and $ c = a + \sqrt {3} \ Q_{30} $, where the zero in the subscript denotes the calculated magnitudes for the unstrained LMO.
 The magnitudes of these distortions that minimize the ground-state energy clearly depends on the Hamiltonian parameters used in the model. For our typical parameters, we find that  $Q_{20}  \approx 0.32$  \AA\ and 
$Q_{30}  \approx - 0.18$ \AA, which are close to the experimental values of $0.28$  \AA\ and $- 0.10$  \AA, respectively.\cite{elemans} 
Optimization of $Q_1$ for the Hamiltonian yields, trivially, the result that $Q_{10}= -g^\prime/K$, while the cell volume is given by $a^2c \approx (\sqrt{2/3} \ Q_{10} )^3$. So, in essence, the experimental volume of LMO, which corresponds to  $Q_{10} = 4. 85 $ \AA\, determines the ratio of $-g^\prime/K$.

We note  that $c$ and $a$ are independent of the strength of the 
in-plane distortion $Q_2$. Therefore, any strain that changes $c$ and $a$ will directly affect the $Q_1$ and $Q_3$ distortions only, whereas the system will have an optimized $Q_2$ distortion in order to achieve the minimum in the ground state energy. This is consistent with  the experimental studies of bulk LMO under pressure, where it has been found that the Mn-O bond length along the
c-axis
changes linearly with 
the lattice parameter of the LMO unit cell, while the Mn-O bond lengths along the plane 
follow a non-linear behavior.\cite{loa, gaudart}

Now, with uniaxial strain, the lengths of the axes become $a \ (1+\varepsilon_{xx})$ and 
$c \ (1+\varepsilon_{zz})$, respectively. The strain parameters completely determine the magnitudes of the breathing and the octahedral stretching modes, viz.,
\begin{eqnarray}
Q_1 &=& Q_{10}(1 + 2\varepsilon_{xx}/3 + \varepsilon_{zz}/3)\nonumber \\
Q_3 &=& Q_{30}(1+\varepsilon_{zz}) + \sqrt{2}Q_{10}(\varepsilon_{zz} - \varepsilon_{xx})/3.
\label{Q1Q3-strain}
\end{eqnarray}
In addition, the Poisson's ratio $\nu$ relates the two strain components,\cite{poisson1} 
$\varepsilon_{xx} = (2 \nu)^{-1} (\nu-1) \varepsilon_{zz}$,
	 so that there is a single strain parameter $\varepsilon_{zz}$ in the problem. 
	 Rewriting Eq. (\ref{Q1Q3-strain}) in terms of this parameter, we have 
\begin{eqnarray} 
Q_1 &=& Q_{10}(1 + \frac{2\nu-1}{3\nu} \varepsilon_{zz}) \nonumber  \\ 
Q_3 &=& Q_{30}(1+\varepsilon_{zz}) + Q_{10} \frac{\sqrt{2}}{3}  \frac{\nu+1}{2\nu}  \varepsilon_{zz}, 
\label{Q1Q3-nu}
\end{eqnarray}
so that in our model, the strain parameter fixes the values of the $Q_1$ and the $Q_3$ distortions. The magnetic configurations and the in-plane JT distortion $Q_2$ are then determined so as to minimize the total energy.

	The value $\nu=1/2$ corresponds to conservation of volume under strain and the
	measured values for materials are not too far from this. For the manganites, the measured value is $\nu \approx 0.4$,\cite{poisson1, poisson2} which we use in our calculations together with the unstrained distortions $Q_{10} = 4.85$ \AA\ and $Q_{30} = -0.18$ \AA\ as indicated above. 
	For the case of conserved volume ($\nu = 1/2$), the value of $Q_1$ does not change with strain, while for $\nu = 0.4$ and $\varepsilon_{zz} = 5\%$, it changes by less than a percent from $Q_{10}$, while $Q_3$ changes significantly. Furthermore,
	because of the form of the Hamiltonian, Eqs. (\ref{htot}) and (\ref{hjt}), $Q_1$ adds just an elastic energy and shifts the energies of both components of the $e_g$ orbitals uniformly, irrespective of the magnetic structure of the Mn core spins, so that the relative energies of the different magnetic structures for a given strain $\varepsilon_{zz}$ are not affected by $Q_1$. 
Thus in essence $Q_3$ remains the sole parameter in the problem for obtaining the ground states for different strains and the two parameters $\varepsilon_{zz}$ and $Q_3$ can be used interchangeably to describe the strain condition. It has been recently shown that for manganites, the strain parameter $\varepsilon_{zz}$ can be varied by as much as $\pm 5\%$ by using different materials as substrates.\cite{poisson1}

\section{Results and Discussions}

{\it Isolated octahedra under strain} -- To begin on familiar grounds, we consider the situation where the intersite hopping
integral $V_{\sigma}$ = 0 in the Hamiltonian (\ref{htot}). This is equivalent to a collection of isolated MnO$_6$ JT centers, each under the applied strain condition. 
The octahedra interact only via the super exchange interaction of the Mn core spins that results in the  Ne\'{e}l AFM order, but the orbital occupancy and the JT distortions are determined by the interaction within a single octahedron.
We see from Fig. \ref{magneticphase} that in this case the magnetic ordering changes from type G AFM to type G$^{\prime}$ AFM beyond a certain value of tensile strain. This can be easily understood as explained below.

\begin{figure}
\includegraphics[width=6cm]{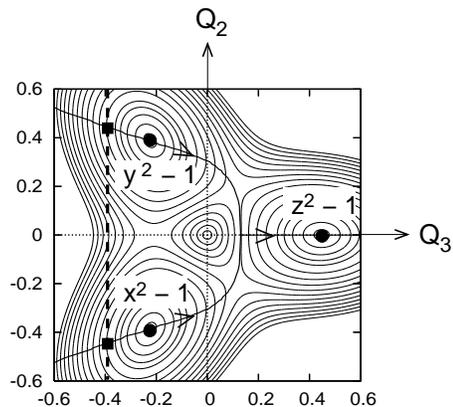}
\caption{\label{gafmcontour} Energy contours of an isolated Jahn-Teller center on the $Q_2-Q_3$ plane with linear and quadratic coupling. The three global minima are indicated by the circular dots along with the corresponding occupied orbital; one of the three minima is occupied in the isolated octahedron. For the octahedron in the solid, the strain parameter $\varepsilon_{zz}$ fixes the magnitude of $Q_3$ and the octahedron occupies the state marked by the square dots.  As $Q_3$ is changed, the system traverses along the lines marked by the arrows.
Beyond a critical value of the tensile strain (i.e., a critical $Q_3$), the orbital order abruptly switches from $x^2-1$ or $y^2-1$ like to $z^2-1$, making thereby a transition
from G to G$^\prime$ state. }
\end{figure}

The energy of an isolated octahedron, with one $e_g$ electron occupied, is well known.\cite{bersuker}
The expression is
\begin{equation}
E = \frac{1}{2}KQ^2 - Q(g^2 + G^2Q^2 + 2gG\cos(3\phi))^{\frac{1}{2}},
\end{equation}
where
$
Q =  \sqrt{Q_2^2 + Q_3^2}$  and      $\phi = \tan^{-1}(Q_2/Q_3).$ This leads to the energy contours
with three equivalent minima at $\phi = 0, \pm 2\pi/3$ with $Q = g/(K-2G)$ and the corresponding occupied orbital as shown in 
Fig. \ref{gafmcontour}.  

Now, when the octahedron is placed in LMO, the strain parameter $\varepsilon_{zz}$ fixes the magnitude of $Q_3$ as given by Eq. (\ref{Q1Q3-nu}), with the result that the state of the octahedron is given by minimizing the energy with respect to $Q_2$ only. The constrained minima for a certain value of $Q_3$ are shown by the square dots in
Fig.  \ref{gafmcontour}. For the isolated octahedron, there are two equivalent minima with either the $x^2-1$ or the $y^2-1$ occupation; In the solid, they become staggered resulting in the checkerboard orbital order in the $ab$ plane.   As strain is changed, $Q_3$ changes and the system traverses along the lines marked by the arrows in the figure. Beyond a certain tensile strain, the strength of the $Q_2$ mode sharply reduces to zero, which is also accompanied by a transition in the orbital occupancy from $x^2 - 1$ or $y^2 - 1$ to $z^2 - 1$. The octahedra interact only via the super exchange term when the hopping term $V_\sigma$ is zero, so that the Ne\'{e}l order is retained throughout, resulting in the transition from type G to G$^\prime$ configuration as the orbital state changes.

{\it Inter-octahedral hopping allowed } -- When the inter-octahedral hopping is allowed ($V_\sigma \neq 0$), it leads to a cooperative JT ordering, i.e., that the JT distortions of the neighboring octahedra are organized so as to
optimize the various energy terms in the solid. The ground-state configuration is determined by a competition between (i) the electron kinetic energy, which would prefer no JT distortion and a FM alignment of the core spins, so that all $e_g$ orbitals are aligned and the kinetic energy gain due to hopping is maximized, (ii) the on-site JT interaction that would prefer a large octahedral distortion, at the same time balancing the elastic energy cost, and (iii) the super exchange energy that prefers antiferromagnetic alignment between all neighboring spins. If the electronic hopping is taken as zero, there is no conflict between the last two terms and a Ne\'{e}l state (G or G$^\prime$) results as discussed above, simultaneously maximizing the JT energy gain and as well as the super exchange term.

\begin{figure}
\includegraphics[width=9cm]{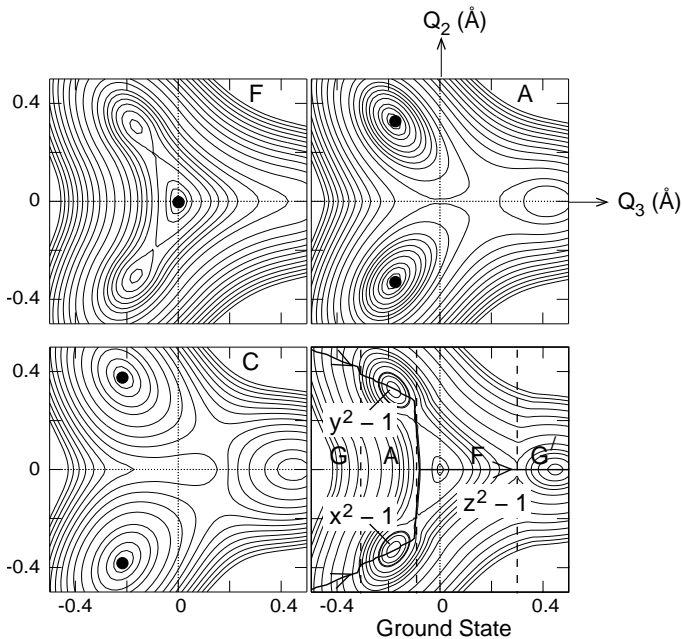}
\caption{\label{contour} Energy contours for different phases and for the minimum-energy phase (bottom right). For a fixed strain (fixed $Q_3$), the global minimum is obtained by varying $Q_2$ for each phase and comparing the minima thus obtained for each phase. The line with arrows indicates the state of the system as strain is varied. As strain changes from compressive to tensile, the magnetic state changes from G $\rightarrow $A $\rightarrow $F $\rightarrow $ G$^\prime$, the magnitude of $Q_2$ slowly diminishes eventually becoming zero, and  finally the orbital order changes from $x^2-1$/$y^2-1$ to $z^2-1$ type.
Contours are obtained by solving Hamiltonian Eq. (\ref{htot}) with the hopping parameter V$_{\sigma}$ = -0.5 eV.}
\end{figure}

With the switching on of the hopping term, all three energies cannot be simultaneously optimized. The resulting competition leads to a staggered orbital order accompanied by a staggered JT distortion, where both $Q_2$ and $Q_3$ distortions occur at each site, with Q$_2$ changing sign  on neighboring sites and the $x^2-1$ and $y^2-1$ orbitals alternately occupied. The system thus stabilizes in the orbital ordered type A AFM state\cite{popovic}.

When strain is applied, other magnetic phases compete, resulting in different ground-state structures as shown in Fig. \ref{magneticphase}.
To gain insight into the possible phases in strained LMO, we have kept the staggered JT distortion and examined four possible magnetic ordering, viz., A, F, G, and C. 
In Fig. \ref{contour} we have shown the energy contours in the $Q_3 - Q_2$ plane for the type F, type A, and type C magnetic ordering. Results for type G ordering remains the same as in Fig. \ref{contour}, irrespective of the magnitude of the hopping integral, since the Anderson-Hasegawa cos ($\theta/2$) factor makes the effective hopping to be zero
anyway. 
For the ferromagnetic ordering, the $e_g - e_g$ hopping is allowed in all the directions. It is the strongest when there is no JT splitting of the $e_g$ orbitals. Therefore the global minimum for the ferromagnetic ordering occurs at the center of the $Q_3 - Q_2$ plane as can seen from the energy contours of Fig. \ref{contour}. As we deviate from the center, JT-splitting reduces the effect of the hopping due to the energy denominator, which therefore weakens the FM stability.

For the type 
A AFM ordering, the $e_g - e_g$ orbital hopping is forbidden 
along the c-axis. Therefore, throughout the $Q_3 - Q_2$ plane, the net gain in the band energy is weaker in
comparison to the FM case. In this case, the competition between the JT energy and band energy lead to two global minima located at $Q_3$ = -0.18 {\AA} and $Q_2$ = $\pm$ 0.32{\AA} (see Fig. \ref{contour} (top right)) which is close to earlier experimental and theoretical studies\cite{loa,popovic,singh}. In fact, taking the super exchange into account, these minima have the lowest energy in the $Q_3 - Q_2$ plane when all the magnetic configurations are considered. 

In the case of type C magnetic ordering, where the antiferromagnetic $ab$ planes are stacked ferromagnetically along the $c$-axis, the hopping between the $z^2 - 1$ orbitals along the $c$-axis is only allowed  and the remaining hopping interactions are forbidden either due to orbital symmetry along the $c$-axis or due to the Anderson-Hasegawa $\cos(\theta/2)$ factor along the plane. As a result, the system gains very little band energy, making this phase unfavorable for all strains.

\begin{figure}
\includegraphics[width=7cm]{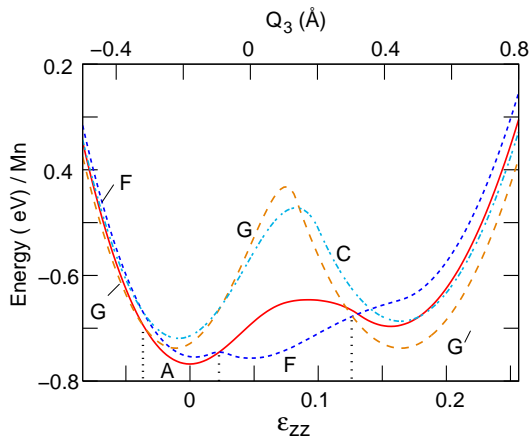}
\caption{(Color online) Total energy for the different phases as a function of strain. Parameters are the same as in Fig. \ref{contour}.
}
\label{totalenergy} 
\end{figure}

{\it Total energy} -- The total energies of the various structures as a function of strain are plotted in Fig. \ref{totalenergy}. It is easy to understand several general features of the energy. First, note that with the energy minimum occurring at zero strain condition ($\varepsilon_{zz}=0)$, the total energy is internally consistent. If the minimum occurred somewhere else, that would indicate that LMO would automatically occur with a different strain condition. The energy of the F structure does approach the minimum of the A structure at $\varepsilon_{zz}=0$, but does not come below it. 

A second feature of the total energy is that for large strains, the octahedral distortion is strong, which leads to a large on-site splitting of the $e_g$ states,
\begin{equation}
\epsilon_\pm = g^\prime Q_1 \pm g \sqrt {Q_2^2+Q_3^2}, 
\label{JT-split}
\end{equation}
the lower orbital being fully occupied and the higher orbital empty. This in turn leads to a  vanishing gain in energy due to hopping between the occupied and the unoccupied states because of the large energy denominator in the second order perturbation theory, irrespective of the magnetic structure and the corresponding Anderson-Hasegawa $\cos (\theta/2) $ factors. The net result is that for large strains, the energies for all structures simply tend to a sum of the elastic energy and the on-site JT energy: 
\begin{equation}
E=(1/2) K Q^2 + g^\prime Q_1 - g \sqrt {Q_2^2+Q_3^2}.
\label{parabola}
\end{equation} 
%

The quadratic dependence on the JT distortion or, equivalently, the strain for large strain for all magnetic structures is clearly seen in Fig.
\ref{totalenergy}. The curves are parabolas with shifted minima for the G and G$^\prime$ structures, where the expression (\ref{parabola}) is exact, since there is no hopping for these structures. For the remaining structures, there is a gain of energy due to hopping, being maximum for the ferromagnetic F structure, where hopping is allowed in all directions in the lattice. This leads to the stability of the F structure for small strain parameters as seen from the figure. The argument is consistent with the fact that the energy gain for small strain progressively increases from C to A to F (as seen in Fig. \ref{totalenergy}), where hopping is allowed along one, two, and three directions, respectively, for these structures.

Another point regarding the total energy of Fig. \ref{totalenergy} is that the ferromagnetic phase is always close by in energy for small strain situations,
which is suggestive of the reason why LMO thin films are usually observed to occur in the ferromagnetic phase, rather than the  A phase of the bulk. 


Finally, one might ask how is it  possible that at larger strains the N\'{e}el  ordered G or G$^\prime$ state wins over the ferromagnetic F state. The reason is that as strain is increased, the progressively larger JT distortions split the $e_g$ states more and more (Eq. \ref{JT-split}).  The kinetic energy gain in the FM phase due to hopping between the occupied e$_g$ state and the neighboring unoccupied e$_g$ states progressively diminishes, due to the energy denominator in the second-order perturbation theory. This can not overcome the super exchange energy cost, so that the N\'{e}el ordered states win in the limit of large strains.

\begin{figure}
\includegraphics[width=7cm]{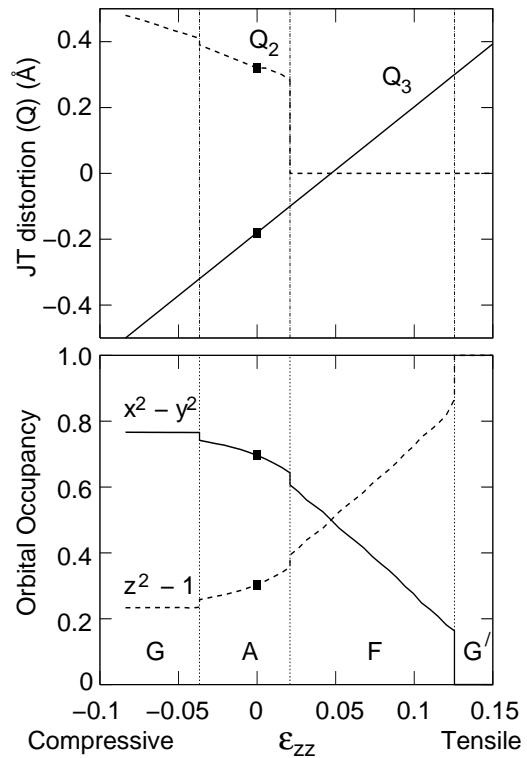}
\caption{\label{occupancy} Variation of the JT distortions (upper panel)
and the occupancy of the Mn e$_g$ orbitals (lower panel) as a function of the
applied strain. 
In our model, $Q_3$ and 
$\varepsilon_{zz}$ are linearly related by Eq. (\ref{Q1Q3-nu}).
 The hopping parameter is
V$_{\sigma}$ = -0.5 eV. Dots denote the unstrained structure.
}
\end{figure}

{\it JT distortions and orbital occupancy} -- 
Turning now to the question of JT distortion strengths, Fig. \ref{occupancy} shows the variation of
$Q_2$ and $Q_3$ modes and the change in the orbital occupancy  as a function of strain. As discussed earlier, the $Q_3$ mode is directly related linearly to $\varepsilon_{zz}$ through Eq. ( \ref{Q1Q3-nu}). Quite interestingly, we find that the magnitude of the in-plane distortion $Q_2$ at first diminishes and then sharply drops to zero as the strain is changed from compressive to tensile. This is due to 
the fact that in the compressive strain condition, the system stabilizes in the type G AFM ordering where the $Q_2$ mode is more favorable (see Fig. \ref{gafmcontour}). On the other hand, in the tensile 
strain condition, in the stable ferromagnetic (F) ordering, kinetic energy gain is maximum if the $e_g$ orbitals are not split, so that  the JT distortion is small or non-existent. 

Strain has a strong effect on the $e_g$ orbital occupancies as indicated in Fig. \ref{contour}. The calculated orbital occupancies are plotted in Fig. \ref{occupancy}. In the A phase, the $x^2-y^2$ and the
$z^2$-1 orbitals add up with the proper linear combination to yield the checkerboard $x^2 - 1$/$y^2 - 1$ orbital ordering on alternate Mn atoms in the $ab$ plane.
 As the strain is increased, the $z^2-1$ orbital becomes more and more occupied with the occupancy becoming one in the G$^\prime$ phase. 
This is because in this limit $Q_3$ is large and $Q_2$ is zero, so that the elongated Mn-O bond along the c-axis results in a strong octahedral crystal field that increases the energy of $x^2-y^2$ orbital, simultaneously lowering that of  $z^2 - 1$ orbital and fully occupying it in the process. This physics is of course incorporated in the linear Jahn-Teller coupling term in our model Hamiltonian Eq. (\ref{q1q2q3}). 
\begin{figure}
\includegraphics[width=5cm]{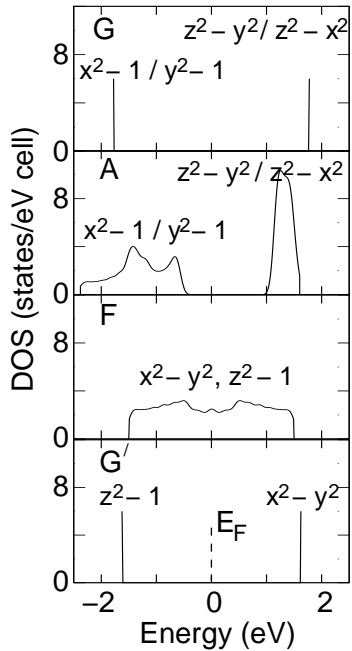}  
\caption{\label{dos}
 Densities of states (DOS)  corresponding to the four strain conditions of Fig. \ref{orbital}, with hopping parameter 
V$_{\sigma}$ = -0.5 eV. In the Ne\'{e}l ordered G and G$^\prime$
phases, complete suppression of hopping leads to the $\delta$-functions in the DOS; In the G phase, alternate Mn sites have the $x^2-1$ or the $y^2-1$ orbitals occupied with the remaining component of the $e_g$ manifold empty, leading to the checkerboard orbital ordering, while in the G$^\prime$ phase, all sites have the $z^2-1$ orbital occupied. In the A phase, allowed hopping in the plane leads to a broadening of the band, but still a gap in the DOS remains. In the ferromagnetic F phase, hopping both in and out of the plane is strong enough to close the gap, leading to a metallic state due to band gap closure. The Fermi energy is taken as the zero of energy in this plot.
}
\end{figure}
%
\begin{figure}
\includegraphics[width=8cm]{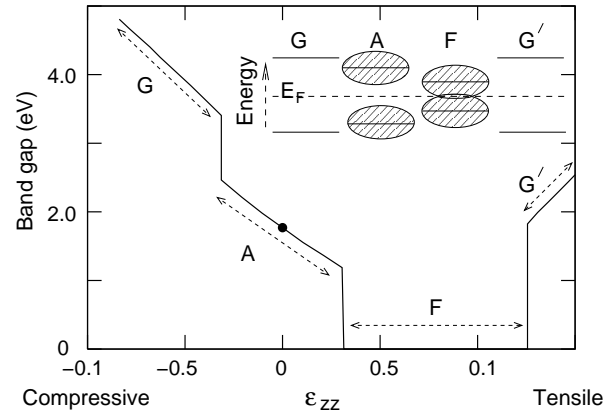}
\caption{\label{bandgap}
Variation of the band gap with strain. Inset shows the typical band structure for the four phases. The black dot indicates the band gap for the unstrained LMO. }
\end{figure}

Beside the magnetic transitions, the other important feature in the strain induced LMO is a metal-insulator transition due to band gap closure. Fig. \ref{dos} shows the densities of states for the various phases, which faithfully indicates the strong JT splitting of the on-site energies of the $e_g$ orbitals, $\Delta  = g \sqrt{Q_2^2+Q_3^2}$, with only a large hopping in the ferromagnetic phase being able to close the gap and produce a metallic phase. 
In Fig. \ref{bandgap}, we have shown the calculated band gap as a function of strain. 
We see that the gap changes abruptly as new phases are entered. We note that for the unstrained case, the magnitude of the band gap obtained with our parameters is 1.76 eV as compared to the measured value of 1.8 eV, obtained from the
photoemission data\cite{saitoh}. This suggests that the experimental band gap under strain may also show changes similar to what is indicated from our model.

\section{Summary}
In summary, we studied the effect of uniaxial strain on the ground-state
structure of LaMnO$_3$ from a model Hamiltonian that included the 
key interactions in the system, viz., the super exchange, the double exchange, and the
Jahn-Teller electron-lattice coupling. Our results reveal the existence of various phases
under varying strain conditions, where the orbital order, magnetic structure, octahedral JT distortions, as well as the conduction properties change under applied strain.
Only the ferromagnetic phase, stable with a small tensile strain in our model, was found to be metallic, while all other phases were found to be insulating. The analysis presented here should be helpful in the interpretation of experiments on epitaxially grown
LaMnO$_3$ 
heterostructures, which are often in the  uniaxial strain condition. 

This work was supported by the U. S. Department of Energy through Grant No.
DE-FG02-00ER45818.

\end{document}